# SUSY-inspired one-dimensional transformation optics

Mohammad-Ali Miri,* Matthias Heinrich, and Demetrios N. Christodoulides

CREOL, College of Optics and Photonics, University of Central Florida, Orlando, Florida 32816–2700, USA
* miri@knights.ucf.edu

**Transformation optics aims to identify artificial materials and structures with desired electromagnetic properties by means of pertinent coordinate transformations. In general, such schemes are meant to appropriately tailor the constitutive parameters of metamaterials in order to control the trajectory of light in two and three dimensions. Here we introduce a new class of one-dimensional optical transformations that exploits the mathematical framework of supersymmetry (SUSY). This systematic approach can be utilized to synthesize photonic configurations with identical reflection and transmission characteristics, down to the phase, for all incident angles, thus rendering them perfectly indistinguishable to an external observer. Along these lines, low-contrast dielectric arrangements can be designed to fully mimic the behavior of a given high-contrast structure that would have been otherwise beyond the reach of available materials and existing fabrication techniques. Similar strategies can also be adopted to replace negative-permittivity domains, thus averting unwanted optical losses.**

**Introduction.** The problem of reconstructing the shape of a potential distribution solely from information carried by its scattering pattern has a long-standing history in a number and diverse disciplines of science and technology [1-5]. Naturally, in such inverse problems, the question of uniqueness is of crucial importance: Are the properties of an object fully determined by its corresponding far-field scattering data? In general, the answer is no. In quantum mechanical settings, for example, one can always identify an N-parameter family of different potentials that support the same discrete set of N bound-state eigenvalues and exhibit similar scattering characteristics [4]. Closely related to this subject is the idea of supersymmetry (SUSY) [6-10]. This mathematical framework emerged in quantum field theory as a means to treat fermions and bosons on equal footing [9]. Subsequently, notions of supersymmetry were utilized to obtain isospectral and phase-equivalent potentials within the context of non-relativistic quantum mechanics [10].

On the other hand, recent developments in transformation optics and optical conformal mapping have brought about novel methodologies to address inverse problems [11-16]. By virtue of coordinate transformations, such schemes can in principle provide the spatial distribution of electric permittivities and magnetic permeabilities that would perform a desired task such as cloaking [17-22]. As one would expect, the material properties required to implement such configurations might not always be available in practice. Clearly of interest would be to develop alternative strategies that allow one to judiciously control the scattering properties of an object, while at the same time reducing the complexity of the structures involved.

As we will see, supersymmetry can provide a new avenue for one-dimensional transformation optics that would have been otherwise impossible using other multi-dimensional approaches (see Fig. 1). Along these lines we introduce appropriate optical transformations in 1D refractive index landscapes and explore their implications in terms of their far-field response. In addition to finding superpartners with similar scattering behavior, systematic SUSY deformations allow us to design systems that exhibit identical complex reflection and transmission coefficients for all incident angles. As a result, two such dielectric objects, however dissimilar, become virtually indistinguishable. Remarkably, the proposed formalism can be employed to synthesize photonic configurations that behave exactly the same way as high refractive index contrast devices – by only utilizing low-contrast dielectric media. Similar methodologies can be employed to substitute negative-permittivity inclusions with purely dielectric media as a means to obtain the intended functionality without introducing any additional loss.

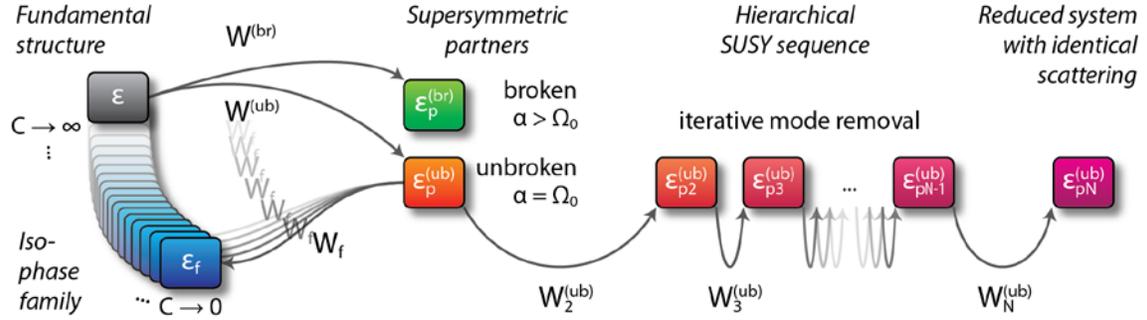

**Fig. 1.** Schematic overview of the different SUSY optical transformations. Starting from a given fundamental structure $\epsilon$, supersymmetric partners $\epsilon_p$ can be constructed. Whereas the broken SUSY system $\epsilon_p^{(br)}$ preserves all bound modes, unbroken SUSY ($\epsilon_p^{(ub)}$) removes the fundamental mode. Regardless, in both cases the intensity reflection and transmission coefficients of the superpartners are identical to those of the fundamental system. In order to maintain the full complex scattering characteristics, a family $\epsilon_f$ of iso-phase structures can be synthesized. Finally, a hierarchical sequence of higher-order superpartners $\epsilon_{p,2...N}^{(ub)}$ may be utilized to obtain a scattering-equivalent structure, which requires a substantially lower refractive index contrast than that involved in the original system $\epsilon$.

**Supersymmetric optical transformations.** In one-dimensional inhomogeneous settings, the propagation of TE polarized waves is known to obey the Helmholtz equation [23] $[\partial_{xx} + \partial_{yy} + k_0^2\epsilon(x)]E_z(x,y) = 0$ where $k_0$ is the vacuum wavenumber and $\epsilon(x)$ is the relative permittivity of a given (fundamental) structure to be emulated via SUSY transformations (see Fig.1). The analysis of TM waves can be carried out in a similar manner (see the supplementary information). In the TE case, the spatial dependence of the electric field $E_z$ can be described via $E_z(x,y) = \psi(x)e^{i\beta y}$. Here, $\beta = k_0 n_0 \sin\theta$ represents the y-component of the wave vector for an incidence angle $\theta$, and $n_0 = \sqrt{\epsilon(x \to \pm\infty)}$ is the background refractive index. By employing the normalized quantities $X = k_0 x$, $Y = k_0 y$ and $\Omega = \beta^2/k_0^2$, the Helmholtz equation then reduces to a 1D Schrödinger-like equation

$$H\psi(X) = \Omega\psi(X). \qquad (1)$$

The resulting Hamiltonian $H = \partial_{XX} + \epsilon(X)$ can be factorized as $H = BA + \alpha$ where the operators $A$ and $B$ are defined as $A = \partial_X + W(X)$, $B = \partial_X - W(X)$, and $\alpha$ is an auxiliary constant of the problem. Here $B = -A^\dagger$, where "†" represents the Hermitian conjugate. The superpotential $W$ can then be obtained as a solution of the Riccati equation [7]:

$$\epsilon(X) = +W' - W^2 + \alpha, \qquad (2)$$

in terms of the fundamental permittivity profile $\epsilon(X)$. Once $W$ has been determined, one can establish a partner Hamiltonian $H_p = AB + \alpha$, which corresponds to a new distribution in the electric permittivity

$$\epsilon_p(X) = -W' - W^2 + \alpha. \qquad (3)$$

As a direct consequence of this construction, the modes $\psi_p$ of the partner potential $\epsilon_p$ are related [24] to the ones of the fundamental through the following expressions $\psi_p \propto (\partial_X + W)\psi$ and $\psi \propto (\partial_X - W)\psi_p$. These latter relations hold for guided waves as well as for radiation modes, and each such pair of states is characterized by a common eigenvalue. We note that two options for choosing $\alpha$ exist: (a) Assuming that the structure supports at least one bound state, one may opt to set α equal to the fundamental mode's eigenvalue, i.e., $\alpha = \Omega_0$. (b) The other possibility is to choose $\alpha > \Omega_0$, irrespective of whether the system supports bound states or not. The first case corresponds to an unbroken SUSY: The two potentials share the guided wave eigenvalue spectra, except for that of the fundamental mode, which does not have a corresponding state in the partner. In the second case, however, SUSY is broken, and the two arrangements share an identical eigenvalue spectrum, including that of the fundamental mode. As an example, Fig. 2(a) depicts the relative permittivity distribution $\epsilon(X) = 1 + \exp[-(X/5)^8]$ corresponding to a step-index-like waveguide; its unbroken and broken SUSY partners are shown in Figs. 2(b,c) respectively. $W$ can also be found analytically [7] via $W = -\partial_X \ln(\psi_0)$ in the unbroken SUSY case, i.e. for $\alpha = \Omega_0$, when $\epsilon$ supports at least one bound state $\psi_0$. In either regime, Eq. (2) can always be solved numerically to obtain the superpotential $W$. An alternative approach is to start with an arbitrary superpotential and construct the two superpartner structures $\epsilon$ and $\epsilon_p$ according to Eqs. (2,3). In this scenario, it still remains to be determined whether SUSY is unbroken or broken. This question can be resolved by the so-called Witten index [6]. In general if $W(X)$ approaches $W_\pm$ at $X \to \pm\infty$, unbroken SUSY requires $W_+ = -W_-$, while a broken SUSY demands that $W_+ = W_-$.

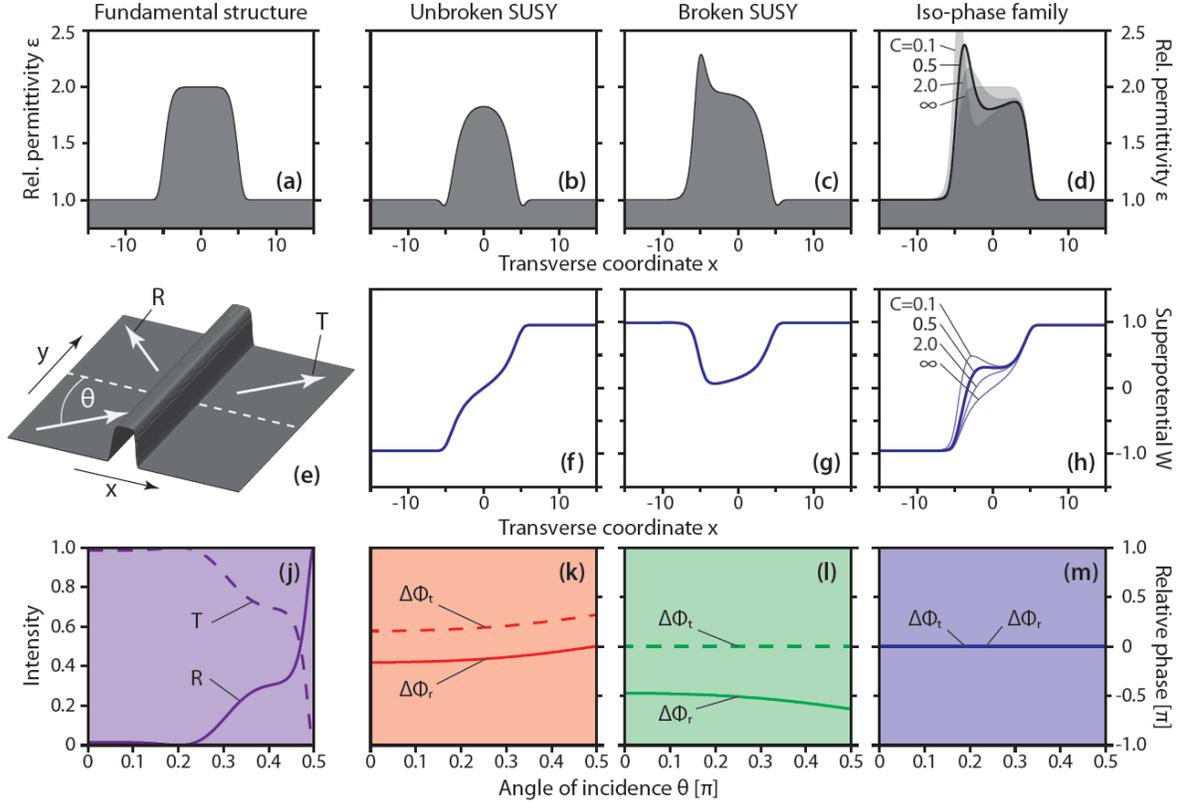

**Fig. 2.** Relative permittivity distributions of the original and the transformed potentials, (a) The fundamental system has a step-like profile $\epsilon(X) = 1 + \exp[-(X/5)^8]$. (b) The superpartner in the unbroken SUSY regime, (c) The superpartner in the broken SUSY case, and (d) phase-equivalent structures. (e) Scattering geometry. (f-h) Superpotentials $W$ corresponding to panels (b-d). (j) Identical reflectivity $R$ (solid line) and transmittivity $T$ (dashed line) corresponding to Figs. 1(a-d). (k-m) Relative phases of the reflection ($\Delta\Phi_r$, solid line) and transmission ($\Delta\Phi_t$, dashed) coefficients of the structures in (b-d) compared to the fundamental system (a) as a function of the incident angle $\theta$. The scattering characteristics were evaluated by means of the differential transfer matrix method [23].

It is important to note that more than one superpotential can exist for any given distribution $\epsilon(X)$. As shown in the supplementary information, one can actually find a parametric family $W_f$ of viable superpotentials that satisfy Eq. (3). Whereas all members of this family lead to the same superpartner $\epsilon_p$, each of them describes a different permittivity distribution $\epsilon$ according to Eq. (2). The resulting parametric family [10,25] of structures $\epsilon_f(X;C)$ is associated with the fundamental distribution $\epsilon$ and its ground state $\psi_0$ as follows:

$$\epsilon_f(X;C) = \epsilon(X) + 2\partial_{XX} \ln\left(C + \int_{-\infty}^{X} \psi_0^2(X')dX'\right), \quad (4)$$

where $C$ represents a free parameter. Figure 2(d) depicts such family members for the fundamental structure $\epsilon$ shown in Fig. 2(a) when $C = 0.1, 0.5$ or $2.0$, respectively. Note that the original permittivity distribution $\epsilon$ is in itself a member of this family, since $\epsilon_f \to \epsilon$ for $C \to +\infty$. All the modes $\psi_f$ of any other member are related to its states $\psi$ according to $\psi_f \propto (\partial_X - W_f)(\partial_X + W)\psi$ (see supplementary information). As a result, all family members share the same guided wave characteristics, e.g. they have identical sets of eigenvalues as in the case of broken SUSY. Figures 2(f-h) provide an overview of the different superpotential functions in the regimes of unbroken and broken SUSY (Figs. 2(b,c)) as well as for the family members shown in Fig. 2(d).

**Scattering characteristics.** Let us now turn our attention to the scattering characteristics of structures connected by SUSY transformations. Consider a plane wave $\exp(iXn_0 \cos\theta + iYn_0 \sin\theta)$ incident from the left, i.e. $X \to -\infty$, as shown schematically in Fig. 2(e). For reasons of simplicity we assume a uniform background medium, $n_+ = n_- = n_0$ at $X \to \pm\infty$; the general case of $n_+ \neq n_-$ is discussed in the supplementary information. Here, the reflected and transmitted waves in the far field are given as $r \exp(-iXn_0 \cos\theta + iYn_0 \sin\theta)$ and $t \exp(iXn_0 \cos\theta + iYn_0 \sin\theta)$ in terms of the complex reflection and transmission coefficients $r$ and $t$ [24]. By adopting similar solutions for the partner scatterer $\epsilon_p$, its respective reflection and transmission coefficients $r_p$ and $t_p$ can readily be found (see Table 1 and supplementary information). Interestingly, the SUSY transformation yields a partner structure with exactly the same absolute values in reflection and transmission, as illustrated in Fig. 2(j). Evidently, all the permittivity distributions from Figs. 2(a-d) display identical reflectivities $R = |r|^2 = |r_p|^2$ and transmittivities $T = 1 - R$ for all angles of incidence. In contrast, the scattering phases depend on whether supersymmetry is broken or not (see Table 1). In the case of unbroken SUSY, both reflection and transmission coefficients acquire additional phases with respect to the fundamental scattering potential. If on the other hand SUSY is broken, the transmission coefficient is the same in both amplitude and phase. Finally, one can show that each member of the parametric family $\epsilon_f$ directly inherits all scattering properties of the original structure, i.e. they are phase-equivalent to $\epsilon$. Figures 2(k-m) illustrate these relations.

**Table 1**. Reflection and transmission coefficients for the different SUSY transformations. $W_- = W(-\infty)$ designates the asymptotic value of the superpotential on the left side of the structure, and $r, t$ are the coefficients of the original structure.

| Coefficient | Unbroken SUSY | Broken SUSY | Iso-phase |
|---|---|---|---|
| Reflection | $r_p = r \cdot \exp\left[-2i \tan^{-1}\left(\frac{n_0 \cos\theta}{W_-}\right)\right]$ | $r_p = r \cdot \exp\left[-2i \tan^{-1}\left(\frac{n_0 \cos\theta}{W_-}\right)\right]$ | $r_f = r$ |
| Transmission | $t_p = t \cdot \exp\left[-2i \tan^{-1}\left(\frac{n_0 \cos\theta}{W_-}\right)\right]$ | $t_p = t$ | $t_f = t$ |

**Wavelength dependence.** So far the performance of these systems has been examined at a given operating wavelength $\lambda_0$. Of importance would be to investigate to what extend their supersymmetric properties persist when the wavelength $\lambda$ varies around $\lambda_0$. As one would expect, even if two dissimilar profiles exhibit the same phases at a given wavelength, their internal light dynamics may gradually undergo different changes with $\lambda$. To elucidate this structural dispersion, we provide the spectral dependence of the difference in transmittivities $\Delta T$ (or reflectivities $\Delta R$) between the fundamental structure (Fig. 2(a)) and its superpartners (Fig. 2(b-d)) as a function of the incidence angle $\theta$, as shown in Figs. 3(a-c). As these figures indicate, this difference only becomes notable in the unbroken SUSY regime (Fig. 3(a)), while it is almost absent under broken SUSY and iso-phase conditions (Figs. 3(b,c)). The difference in the corresponding reflection phases is similarly presented in Figs. 3(d-f). The dashed lines trace the abrupt phase jumps of $\pi$, which mark the resonances in the two partners and intersect at the design wavelength $\lambda_0$. Evidently, the iso-phase design displays the greatest resilience with respect to spectral deviations. Note that resonances play no role in the transmission phases, as can be seen in Figs. 3(g-j). In this latter case, the iso-phase system again proves to be the least susceptible to spectral deviations. These results demonstrate that SUSY transformations can be robust over a broad spectral range around the design wavelength.

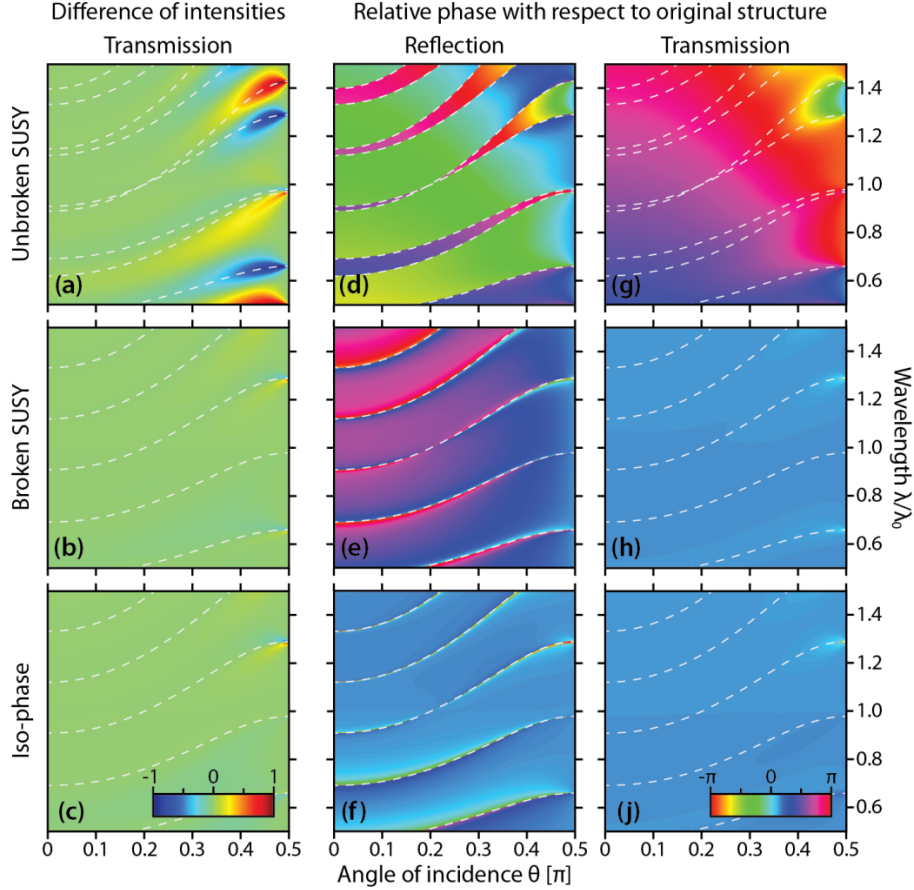

**Fig. 3.** Reflection/transmission characteristics of structures obtained by SUSY transformations depicted in Fig. 2 as function of wavelength $\lambda$ and angle of incidence $\theta$. (a-c) Intensity difference in transmission. (d-f) Relative phases in reflection and (g-j) Relative phases in transmission. The dashed lines follow the resonance-induced $\pi$ phase jumps in fundamental structure and unbroken-SUSY partner. Top row: Unbroken SUSY, Middle row: Broken SUSY, bottom row: Iso-phase case ($C = 0.5$).

**Index-contrast reduction.** One of the main challenges in designing optical systems is the limited dynamic range of refractive indices associated with available materials. This issue becomes particularly acute when high contrast arrangements are desirable. For example, the number of grating unit cells required to achieve a certain diffraction efficiency grows with the inverse logarithm of the index contrast $n_2/n_1$ between the individual layers [23]. As it turns out, SUSY optical transformations can be utilized to reduce the index contrast needed for a given structure. This can be done through a hierarchical ladder of superpartners, i.e. sequentially removing the bound states of the original high-contrast arrangement (Fig. 4a). As a general trend, each successive step demands less contrast in the corresponding index landscape than the previous one (Fig. 4b). The ultimate result is a low-contrast equivalent structure that fully inherits the reflectivity and transmittivity of the original configuration (Figs. 4c,d).

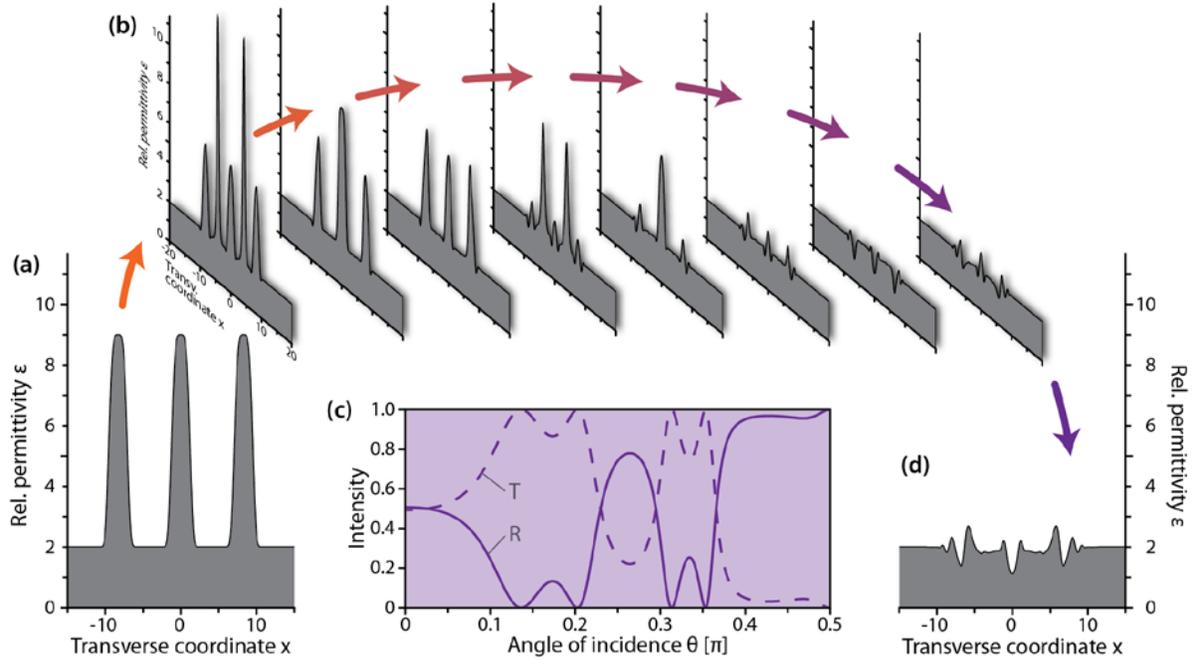

**Fig. 4.** (a) Hypothetical high-contrast dielectric layer arrangement that supports $N = 9$ guided modes. (b) Hierarchical sequence of partner structures obtained through iterative SUSY transformations. (c) Despite the general trend towards lower-contrast configurations, each intermediate step inherits the reflectivity and transmittivity of the fundamental system (a). (d) The resulting low-contrast structure is free of bound states and faithfully mimics the intensity scattering characteristics of the original high-contrast configuration for all angles of incidence.

**Replacing negative-permittivity features**. Finally, SUSY transformations can provide a possible avenue in replacing negative-permittivity inclusions (typically accompanied by losses) by purely dielectric materials. In this respect, inverse SUSY transformations, which now add modes with certain propagation constants to a given structure, can instead be used to locally elevate the permittivity (see supplementary information). Along similar lines, it is possible to find superpotentials that relate a structure with metallic or negative permittivity regions to an equivalent arrangement with entirely positive $\epsilon$, as depicted in Fig. 5. Here we make use of the fact that in a broken-SUSY transformation, the spatial average of $\epsilon$ happens to be a conserved quantity. Therefore, changes in the broader vicinity of the original metal-dielectric structure can be used to achieve this goal.

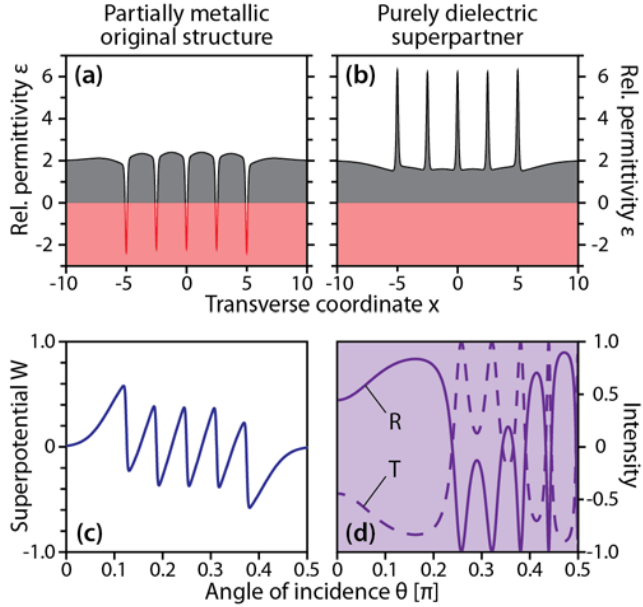

**Fig. 5.** (a) A metal-dielectric grating arrangement comprising five layers of negative electrical permittivity (red sections). (b) An entirely dielectric superpartner grating constructed in the broken SUSY regime, using the respective superpotential (c). (d) Despite the absence of any metallic regions, the equivalent structure exhibits identical reflectivities/transmittivities.

**Discussion.** In conclusion, we have introduced a new type of supersymmetric optical transformations for arbitrary one-dimensional refractive index landscapes. Compared to conventional transformation optics, our approach poses significantly less stringent requirements on the constituent parameters, and does not involve any modifications to the magnetic response of the materials involved. This method can be utilized to construct photonic arrangements that faithfully mimic the behavior of high-index-contrast or even metal-dielectric structures. SUSY transformation optics may have potential applications in a wide range of scenarios that rely on engineered scattering and transmission properties, such as for example optical metasurfaces, anti-reflection coatings and diffraction gratings. Of interest will be to explore how the aforementioned strategies could be paired up with recently developed transformation schemes for guided-wave photonics based on dielectric materials [26,27].

**Acknowledgment**
We acknowledge the financial support from National Science Foundation (NSF) (1128520), and from Air Force Office of Scientific Research (AFOSR) (FA9550-12-1-0148, FA9550-14-1-0037). M.H. was supported by the German National Academy of Sciences Leopoldina (LPDS 2012-01).

# Supplementary information for

# SUSY-inspired one-dimensional transformation optics

Mohammad-Ali Miri, Matthias Heinrich, and Demetrios N. Christodoulides

**S1. Iso-phase potential families**

For any given distribution of the relative electrical permittivity $\epsilon(X)$, an infinite number of viable superpotentials exist. To show this, let us start from Eq. (3), which relates the superpartner $\epsilon_p$ to the superpotential $W$. Starting from a particular $W$, this solution can be generalized by adopting the form $W_f = W + 1/v$, in which case the unknown function $v$ satisfies $(\partial_X - 2W)v = 1$. Direct integration readily leads to $v = e^{+2\int W dX}(C + \int e^{-2\int W dX} dX)$, where $C$ is an arbitrary constant, giving rise to a parametric family $W_f$ of superpotentials

$$W_f(X;C) = W + \partial_X \ln(C + \int e^{-2\int W dX} dX). \qquad (S1)$$

Whereas all members of this family correspond to the same superpartner $\epsilon_p$, each of them describes a different distribution according to Eq. (2). The resulting family of structures $\epsilon_f$ is related to the fundamental one $\epsilon$ according to

$$\epsilon_f(X;C) = \epsilon(X) + 2\,\partial_{XX} \ln(C + \int e^{-2\int W dX} dX), \qquad (S2)$$

If the superpotential $W$ has been specifically obtained from the bound state $\psi_0$, one obtains Eq. (4). Note that since $\epsilon_f \to \epsilon$ for $C \to +\infty$, the fundamental structure is itself a member of this family. The modes $\psi_f$ of any other $\epsilon_f$ can be found by transforming the states $\psi$ of $\epsilon$ according to

$$\psi_f \propto (\partial_X - W_f)(\partial_X + W)\psi. \qquad (S3)$$

Based on this relation, which holds for all bound modes as well as scattered states, one can show that the family $\epsilon_f$ of iso-phase potentials exhibit identical reflection- and transmission coefficients, down to the phase, as the fundamental structure $\epsilon$.

**S2. Reflection/transmission coefficients of superpartner structures**

In order to relate the scattering characteristics of a superpartner to that of its associated fundamental structure, let us first consider an incident plane wave described by $\exp(iXn_0 \cos\theta + iYn_0 \sin\theta)$ impinging on both structures from the left side. The reflected and transmitted waves in the fundamental system are then described by $r\exp(-iXn_0 \cos\theta + iYn_0 \sin\theta)$ and $t\exp(iXn_0 \cos\theta + iYn_0 \sin\theta)$, respectively. Accordingly, the corresponding waves in the superpartner geometry are given by $r_p \exp(-iXn_0 \cos\theta + iYn_0 \sin\theta)$ and $t_p \exp(iXn_0 \cos\theta + iYn_0 \sin\theta)$. By applying the relation $\psi_p \propto (\partial_X + W)\psi$ (that holds between the wave functions of any state pairs in the two structures) to these radiation states, one finds

$$(e^{+iXn_0 \cos\theta} + r_p e^{-iXn_0 \cos\theta}) \propto ((+in_0 \cos\theta + W_-)e^{+iXn_0 \cos\theta} + (-in_0 \cos\theta + W_-)re^{-iXn_0 \cos\theta}), \qquad (S4)$$

for $X \to -\infty$, and

$$t_p e^{+iXn_0 \cos\theta} \propto (+in_0 \cos\theta + W_+) t e^{+iXn_0 \cos\theta}, \quad (S5)$$

for $X \to +\infty$. In these two equations $W_\pm$ denotes the limit of $W$ at $X \to \pm\infty$ respectively. Taking into account that both equations should have the same proportionality constant, one can show that the reflection and transmission coefficients are related via:

$$r_p = \frac{W_- - in_0 \cos\theta}{W_- + in_0 \cos\theta} r, \quad (S6.a)$$

$$t_p = \frac{W_+ + in_0 \cos\theta}{W_- + in_0 \cos\theta} t. \quad (S6.b)$$

Obviously, the reflectivity $R = |r|^2 = |r_p|^2$ as well as the transmittivity $T = 1 - R = |t|^2 = |t_p|^2$ of the superpartner structures is identical. The specific relations for the complex coefficients given in Table 1 in turn follow from $W_+ = -W_-$ in the case of unbroken supersymmetry and $W_+ = W_-$ in the broken supersymmetry regime.

To derive similar expressions for the family of iso-phase structures $\epsilon_f(X; C)$, we label reflected and transmitted waves as $r_f \exp(-iXn_0 \cos\theta + iYn_0 \sin\theta)$ and $t_f \exp(iXn_0 \cos\theta + iYn_0 \sin\theta)$ respectively. One can then relate $r_f$ and $t_f$ to $r$ and $t$ via the transformation $\psi_f \propto (\partial_X - W_f)(\partial_X + W)\psi$. On the other hand, one can write the family of associated superpotentials as

$$W_f(X; C) = W + \partial_X \ln\left(C + \int_{-\infty}^X \psi_0^2(X')dX'\right), \quad (S7)$$

by applying Eq. (5) of the main text and using the fact that $W$ can be written as $W = -\partial_X \ln(\psi_0)$ in terms of the ground state of the fundamental structure. Consequently, $W_f$ and $W$ have the same asymptotic behavior, i.e., $W_{f,\pm} = W_\pm$. In the far field $X \to \pm\infty$, the transformation simplifies to $\psi_f \propto (\partial_X - W)(\partial_X + W)\psi$. Note that $(\partial_X - W)(\partial_X + W) = (\partial_{XX} + W' - W^2) = H - \alpha$, and therefore $\psi_f \propto (H - \alpha)\psi = (\Omega - \alpha)\psi$. Since the scattered waves depend on $Y$ according to $\exp(iYn_0 \sin\theta)$, the corresponding eigenvalue in the Helmholtz equation is given by $\Omega = n_0^2 \sin^2\theta$. Hence, $\psi_f \propto (n_0^2 \sin^2\theta - \alpha)\psi$, and therefore:

$$\left(e^{+iXn_0 \cos\theta} + r_f e^{-iXn_0 \cos\theta}\right) \propto (n_0^2 \sin^2\theta - \alpha)\left(e^{+iXn_0 \cos\theta} + r e^{-iXn_0 \cos\theta}\right), \quad (S8)$$

for $X \to -\infty$, and

$$t_f e^{+iXn_0 \cos\theta} \propto (n_0^2 \sin^2\theta - \alpha) t e^{+iXn_0 \cos\theta}, \quad (S9)$$

for $X \to +\infty$. Given that both equations should be normalized with respect to the same constant, it directly follows that in the iso-phase scenario

$$r_f = r, \quad (S10.a)$$

$$t_f = t. \quad (S10.b)$$

### S3. Configurations with dissimilar backgrounds

Even though the formalism of the main text was developed by assuming that the background medium is the same $(n_- = n_+ = n_0)$, superpartners obeying Eqs. (2,3) can be generated even if $n_- \neq n_+$ (here $n_\pm = \sqrt{\epsilon(\pm\infty)}$ and ). Figure S1(a) shows an example of a step-like distribution in the relative permittivity; the corresponding superpartner $\epsilon_p$ is depicted in Fig. S1(b). The superpotential mediating between them no longer has the same absolute value at $X \to \pm\infty$, but rather is shifted by a certain offset related to the height of the potential step. Neither of the two partner structures does support any guided modes, therefore supersymmetry is necessarily broken. As depicted in Fig. S1(c), numerical results obtained by differential transfer matrix method [S2] show that the angle-dependent power reflectivities $R = |r|^2 = |r_p|^2$ and transmittivities $T = 1 - R = |t|^2 \cdot (n_+/n_-)(\cos\theta_+ / \cos\theta_-) = |t_p|^2 \cdot (n_+/n_-)(\cos\theta_+ / \cos\theta_-)$ remain identical for both systems (Fig. S1(c)), while the phases do not (Fig. S1(d)).

Analytical relations of the scattering coefficients can be found by following analogous steps to those in the previous section. The results are

$$r_p = \frac{W_- - in_-\cos\theta_-}{W_- + in_-\cos\theta_-} r, \qquad \text{(S11.a)}$$

$$t_p = \frac{W_+ + in_+\cos\theta_+}{W_- + in_-\cos\theta_-} t, \qquad \text{(S11.b)}$$

where $\theta_-$ and $\theta_+$ represent the incident and transmitted wave angles related through Snell's law $n_-\sin\theta_- = n_+\sin\theta_+$. It follows that $|r|^2 = |r_p|^2$ and $|t|^2 = |t_p|^2$.

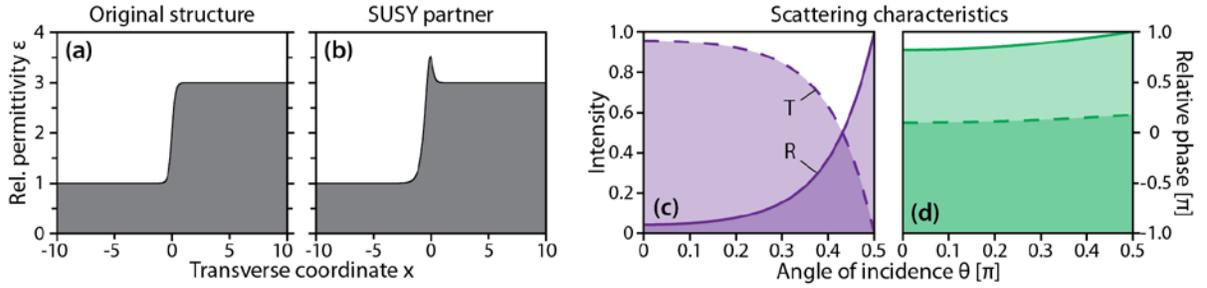

**Fig. S1.** (a) Step-like potential $\epsilon(X) = 2 + \tanh(X/0.3)$ and (b) its superpartner $\epsilon_p(X)$. (c) Despite the potential offset, the intensities of reflected and transmitted waves are identical for both structures, (d) but the phases remain different as expected for the broken SUSY regime.

## S4. Inverse SUSY transformation

In the unbroken symmetry regime, the conventional SUSY transformation may remove a mode from a given fundamental structure $\epsilon$. In doing so, the total area of the relative permittivity is reduced (see Fig. 2(a,b)). This can be shown easily by noting that in the unbroken supersymmetry regime the two superpartners are related via

$$\epsilon_p - \epsilon = -2W'. \qquad \text{(S12)}$$

After integrating both sides of this equation we get

$$\int_{-\infty}^{+\infty} \epsilon_p(X)dX - \int_{-\infty}^{+\infty} \epsilon(X)dX = -2(W_+ - W_-). \qquad \text{(S13)}$$

Since in the unbroken supersymmetry regime $W_- \neq W_+$, the SUSY transformation cannot preserve the total area of the relative permittivity distribution.

On the other hand, one can utilize an inverse SUSY transformation and add a bound state to a given structure $\epsilon$, and in doing so elevate the total area of a given permittivity distribution. We factorize the fundamental Hamiltonian as $H = AB + \alpha$ and define the partner Hamiltonian as $H_{\text{elev}} = BA + \alpha$. Consequently, the two superpartner permittivity distributions can be written as:

$$\epsilon(X) = -W'_{\text{elev}} - W^2_{\text{elev}} + \alpha, \qquad \text{(S14.a)}$$

$$\epsilon_{\text{elev}}(X) = +W'_{\text{elev}} - W^2_{\text{elev}} + \alpha. \qquad \text{(S14.b)}$$

Equation (S14.a) can be solved numerically to obtain the superpotential $W_{elev}$, and from that the partner structure $\epsilon_{elev}$ can be constructed through Eq. (S14.b). As expected in this case the partner structure $\epsilon_{elev}$ exhibits all the guided mode eigenvalue spectrum of the fundamental structure $\epsilon$, as well as an additional guided mode, which takes the place of its previous ground state. As it turns out, the eigenvalue of this additional state is given by the factorization parameter $\alpha$. Note that any value $\alpha > \Omega_0$ can be chosen, where $\Omega_0$ represents the ground state eigenvalue of the fundamental structure. Figure S2 illustrates the inverse SUSY transformation for the specific example of the fundamental structure discussed in Fig. 2 of the main text.

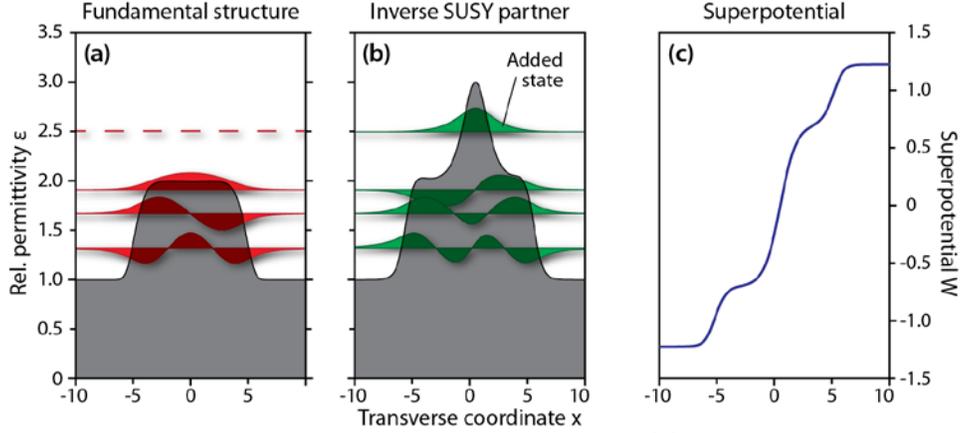

**Fig. S2.** Inverse SUSY transformation. (a) Step-index-like fundamental structure $\epsilon(X)$ supporting $N = 3$ guided modes. (b) Its inverse superpartner $\epsilon_{elev}(X)$, supporting a phase matched mode associated with each of the fundamental structure's states as well as an additional with the desired eigenvalue $\alpha = 2.5$. (c) Superpotential relating the two structures.

**S5. SUSY formalism for the TM polarization**

Under TM polarization conditions the magnetic field component satisfies the equation:

$$\left(\partial_{xx} + \partial_{yy} - \partial_x(\ln \epsilon)\partial_x + k_0^2 \epsilon(x)\right) H_z(x, y) = 0. \quad (S15)$$

In this case again by considering stationary solutions of the form $H_z = \psi(x)e^{i\beta z}$ and after using normalized coordinates $X = k_0 x, Y = k_0 y$ and assuming $\Omega = \beta^2/k_0^2$ we reach at:

$$(\partial_{XX} - \partial_X(\ln \epsilon)\partial_X + \epsilon)\psi = \Omega \psi. \quad (S16)$$

As is, this is not a Schrödinger-like equation, and hence the factorization technique cannot be directly applied. On the other hand, by using the transformation $\psi(X) = \sqrt{\epsilon} f(X)$, this equation can be converted to the desired form:

$$(\partial_{XX} + V_{eff})f = \Omega f, \quad (S17)$$

where $V_{eff}$ is an effective potential that can be expressed in terms of the relative permittivity $\epsilon$ as $V_{eff} = \epsilon - \frac{3}{4}(\epsilon'/\epsilon)^2 + \frac{1}{2}(\epsilon''/\epsilon)$. This relation can also be rewritten as:

$$V_{eff} = \epsilon + \left(\frac{\epsilon'}{2\epsilon}\right)' - \left(\frac{\epsilon'}{2\epsilon}\right)^2. \quad (S18)$$

Following the SUSY formalism, the two superpartner effective potentials can now be written in terms of the superpotential $W$ via

$$V_{eff}(X) = +W' - W^2 + \alpha, \quad (S19.a)$$

$$V_{\text{eff},p}(X) = -W' - W^2 + \alpha. \qquad \text{(S19.b)}$$

One can then reconstruct the relative permittivity of the partner structure $\epsilon_p$ from its corresponding effective potential $V_{\text{eff},p}$ by numerically solving the nonlinear equation

$$V_{\text{eff},p} = \epsilon_p + \left(\frac{\epsilon_p'}{2\epsilon_p}\right)' - \left(\frac{\epsilon_p'}{2\epsilon_p}\right)^2. \qquad \text{(S20)}$$

Of importance will be to show that these transformations preserve the reflection/transmission properties of SUSY partners in the TM case. For this reason note that for effective potentials described in Eqs. (S19), we have the intervening relations $f_p \propto (\partial_X + W)f$ and $f \propto (\partial_X - W)f_p$ therefore:

$$\psi_p/\sqrt{\epsilon_p} \propto (\partial_X + W)\psi/\sqrt{\epsilon}, \qquad \text{(S21.a)}$$

$$\psi/\sqrt{\epsilon} \propto (\partial_X - W)\psi_p/\sqrt{\epsilon_p}. \qquad \text{(S21.b)}$$

Obviously presence of the relative permittivities in the denominator makes these last equations different than those of the TE case. However for calculating the reflection/transmission coefficients (see Supplementary section S2) only the asymptotic behavior of these equations at $X \to \pm\infty$ are of our interest. On the other hand, according to Eqs. (S18–S20), the SUSY partners have the same asymptotic behaviors. Therefore one can use the exact same analysis of the Supplementary section S2 based on Eqs. (S21) to show that the reflection/transmission coefficients of SUSY partners constructed for the TM polarization are related via equations given in Table. 1 of the main text.

**Supplementary references**